\documentclass[a4paper,11pt]{article}
\pdfoutput=1
\usepackage{jheppub}
\usepackage{amsmath}
\newcommand{\Tr}{\mathrm{Tr}\,}
\newcommand{\tr}{\mathrm{tr}\,}

\newcommand{\cF}{{\mathcal{F}}}

\newcommand{\cN}{{\mathcal{N}}}
\newcommand{\cO}{{\mathcal{O}}}

\newcommand{\cW}{{\mathcal{W}}}

\newcommand{\one}{{\rm 1\kern -.9mm l}}

\newcommand{\be}{\begin{equation}}
\newcommand{\ee}{\end{equation}}
\newcommand{\p}{\partial}

\newcommand{\eone}{\epsilon_1}
\newcommand{\etwo}{\epsilon_2}

\newdimen\tableauside\tableauside=1.0ex
\newdimen\tableaurule\tableaurule=0.4pt
\newdimen\tableaustep
\def\phantomhrule#1{\hbox{\vbox to0pt{\hrule height\tableaurule
width#1\vss}}}
\def\phantomvrule#1{\vbox{\hbox to0pt{\vrule width\tableaurule
height#1\hss}}}
\def\sqr{\vbox{%
  \phantomhrule\tableaustep
\hbox{\phantomvrule\tableaustep\kern\tableaustep\phantomvrule\tableaustep}%
  \hbox{\vbox{\phantomhrule\tableauside}\kern-\tableaurule}}}
\def\squares#1{\hbox{\count0=#1\noindent\loop\sqr
  \advance\count0 by-1 \ifnum\count0>0\repeat}}
\def\tableau#1{\vcenter{\offinterlineskip
  \tableaustep=\tableauside\advance\tableaustep by-\tableaurule
  \kern\normallineskip\hbox
    {\kern\normallineskip\vbox
      {\gettableau#1 0 }%
     \kern\normallineskip\kern\tableaurule}%
  \kern\normallineskip\kern\tableaurule}}
\def\gettableau#1 {\ifnum#1=0\let\next=\null\else
  \squares{#1}\let\next=\gettableau\fi\next}
\newcommand{\floor}[1]{\left\lfloor #1 \right\rfloor}

\def\XXint#1#2#3{{\setbox0=\hbox{$#1{#2#3}{\int}$}
     \vcenter{\hbox{$#2#3$}}\kern-.5\wd0}}

\usepackage{tikz}
\usetikzlibrary{graphs}
\usetikzlibrary{trees}
\usetikzlibrary{calc}
\usetikzlibrary{decorations.markings}

\tikzstyle{gauge} = [circle, text centered, draw=black, minimum height=1.2cm]
\tikzstyle{flavor} = [rectangle, text centered, draw=black, minimum height=1.2cm,minimum width=1.2cm]

\tikzstyle{gaugeS} = [circle, text centered, draw=black, minimum height=5ex]
\tikzstyle{flavorS} = [rectangle, text centered, draw=black,minimum height=5ex,minimum width=5ex]

\makeatletter
\pgfdeclareshape{barn}{
    \inheritsavedanchors[from={rectangle}]
    \inheritanchorborder[from={rectangle}] 
    \inheritanchor[from={rectangle}]{center}
    \inheritanchor[from={rectangle}]{south}
    \inheritanchor[from={rectangle}]{west}
    \inheritanchor[from={rectangle}]{east}
    \inheritbackgroundpath[from={rectangle}]
   
    \backgroundpath{
    \southwest \pgf@xa=\pgf@x \pgf@ya=\pgf@y
    \northeast \pgf@xb=\pgf@x \pgf@yb=\pgf@y
    \pgf@xc=\pgf@xb \advance\pgf@xc by-\pgf@xa     
    \pgf@yc=\pgf@yb \advance\pgf@yc by-\pgf@ya    
    \advance\pgf@yb by-0.3\pgf@yc
    \pgfpathmoveto{\pgfpoint{\pgf@xa}{\pgf@ya}}
    \pgfpathlineto{\pgfpoint{\pgf@xa}{\pgf@yb}}
    \pgfpatharc{180}{0}{.5\pgf@xc}  
    \pgfpathlineto{\pgfpoint{\pgf@xb}{\pgf@ya}}
    \pgfpathclose
}
}
\makeatother

\tikzstyle{gaugedflavor} = [barn,draw,text centered, minimum height=1.3cm,minimum width=1.3cm,draw=black]
\tikzstyle{gaugedflavorS} = [barn,draw, text centered, minimum width=6ex,minimum height=6ex,draw=black]

\pgfdeclareshape{cross}{
  \inheritsavedanchors[from={rectangle}]
    \inheritanchorborder[from={rectangle}] 
     \inheritanchor[from={rectangle}]{north}
    \inheritanchor[from={rectangle}]{center}
    \inheritanchor[from={rectangle}]{south}
    \inheritanchor[from={rectangle}]{west}
    \inheritanchor[from={rectangle}]{east}
\inheritbackgroundpath[from={rectangle}]
    \backgroundpath{
    \draw[solid] (.6ex,-.6ex) -- (-.6ex,.6ex);
    \draw[solid] (.6ex,.6ex) -- (-.6ex,-.6ex);
    }
}

\title{\boldmath Chiral rings for surface operators in 4d and 5d SQCD}
\author{Jong-Hyun Baek}

\emailAdd{jonghbaek@gmail.com}

\abstract{Chiral rings of two-dimensional (2,2) theories coupled to 4d $\mathcal{N}=2$ theories with matter hypermultiplets are studied. Specifically, the vacua of the twisted superpotential of the 2d theories with vanishing sum of matter charges are computed by considering the resolvent of the bulk theory. The solutions to the chiral ring equations are also obtained from the instanton partition function via Higgsing for simple surface operators and via the orbifold description for full surface operators. These 2d/4d coupled theories are lifted to 3d/5d theories and vacua are found similarly in two different methods: by solving the 3d chiral ring equations taking into account the effect of 5d resolvent and by computing the 5d instanton partition function in the presence of a surface operator. We also check the Seiberg-like duality for both 2d/4d and 3d/5d coupled systems with a specific Chern-Simons coefficient for the latter.  
}


\begin{document}
\maketitle
\flushbottom

\section{Introduction}
\label{sec:intro}

Surface defects which preserve some of the supercharges in supersymmetric gauge theories have been studied extensively over the years \cite{Gukov:2006jk,Gukov:2008sn,Alday:2009fs, Taki:2009zd, Kozcaz:2010af,Alday:2010vg,Dimofte:2010tz,Maruyoshi:2010iu,Taki:2010bj,Awata:2010bz, Kozcaz:2010yp, Wyllard:2010rp,Marshakov:2010fx,Wyllard:2010vi,Kanno:2011fw,Gaiotto:2013sma,Gomis:2014eya,Nawata:2014nca,Gaiotto:2014ina,Bullimore:2014awa,Gomis:2016ljm,Pan:2016fbl,Ashok:2017odt}. In four dimensional $\cN=2$ theories, half-BPS surface defects can be described either as monodromy defects \cite{Gukov:2006jk,Gukov:2008sn} or as two-dimensional degrees of freedom corresponding to $\cN=(2,2)$ gauged linear sigma models (GLSM) coupled to the bulk theory \cite{Gaiotto:2009fs,Gukov:2014gja}.

In this work, we focus on the two dimensional $(2,2)$ quiver GLSMs with a vanishing sum of matter charges. The flavor symmetries are gauged by the 4d fields, which give twisted masses to the chiral fields in two dimensions. In the Higgs branch, the GLSM flows to a nonlinear sigma model and its target space is a non-compact Calabi-Yau  \cite{Witten:1993yc, Hanany:1997vm}. In the Coulomb branch, after integrating out the matter chiral fields, the GLSM is described by the twisted superpotential, which is a function of the adjoint scalar in the 2d vector multiplet. 

The twisted chiral ring equations are obtained by extremizing the twisted superpotential with respect to the adjoint scalars. The effect of coupling to the 4d theory on the chiral ring equation is accounted by the four dimensional resolvent \cite{Cachazo:2002ry, Gaiotto:2013sma}. Massive vacua can be computed by solving the chiral ring equations around a classical vacuum in series expansion of coupling constants. Evaluating the twisted superpotential at the solution, one obtains the twisted superpotential that depends on both 2d and 4d gauge couplings. It can be compared with the quantity found from the instanton partition function of the bulk theory in the presence of the surface operator. Similar computation was carried out in \cite{Ashok:2017lko} when the 4d theory is $\cN=2$ super Yang-Mills.
Here, we consider surface defects in 4d $\cN=2$ theories with hypermultiplets. 

We compute the instanton partition function in the presence of a surface operator by two methods. For simple surface operators, Higgsing prescription is used \cite{Kozcaz:2010af}. This is equivalent to the geometric transition picture, where a D2 brane appears at special values of the Coulomb branch parameters in 4d $SU(N)\times SU(N)$ theory. It is also related to a degenerate operator insertion in 2d $W_N$ CFT in the context of the AGT relation \cite{Alday:2009fs}. In 6d perspective, these surface operators are M2-branes, which are codimension 4 defects \cite{Gomis:2014eya}.  

The instanton partition function of a full surface operator can be computed by orbifold method \cite{Kozcaz:2010yp, Wyllard:2010vi,Kanno:2011fw}. The 4d gauge group $SU(N)$ is broken by the surface defect to a product of $U(n_I)$'s, where $n_I$'s are $M$ integers, which is a partition of $N$. The instanton partition function of the defect is equivalent to that in the orbifold space, $\mathbb{C}\times\mathbb{C}/\mathbb{Z}_M$ \cite{Braverman:2004vv,Braverman:2004cr,Feigin:2008}. We use the equivariant characters at the fixed points of the tangent space of the moduli space of instantons in the orbifold, which have been known in the literature. From the 6d viewpoint, this surface operator is engineered by an M5-brane as a codimension 2 defect. The relation between codimension 2 and 4 defects is discussed in \cite{Frenkel:2015rda}.   

When the four dimensional theory has hypermultiplets, we find that the codimension 2 and 4 defects couple differently to the four dimensional hypermultiplets. For instance, in 4d $SU(2)$ SYM, a simple surface operator is equivalent to a full surface operator. So it can be viewed either as a codimension 4 or 2 defect. When the 4d $SU(2)$ theory has two fundamentals and two anti-fundamentals, the codimension 4 defect couples to either two fundamentals or two anti-fundamentals, whereas the codimension 2 defect couples symmetrically to one fundamental and one anti-fundamental. This can be seen from the instanton partition functions computed by two different methods. We will consider examples of simple surface operators in $SU(2)$ and $SU(3)$ SCFTs and a full surface operator in $SU(3)$ theory with four hypermultiplets.       

We also consider a 5d lift of the 4d theories by adding a circle. Defects in 5d, which wrap around the circle direction, can be studied similarly. The 3d sigma model is described by the twisted superpotential after all KK contributions are taken into account. The Chern-Simons terms can be present in both 3d and 5d theory but we will consider the cases of zero Chern-Simons levels. Five-dimensional dynamics affects the chiral ring relation through the resolvent of the 5d theory. 

There exists a duality of 2d theories corresponding to the same surface operator \cite{Gorsky:2017hro}. We study the Seiberg-like duality \cite{Seiberg:1994pq,Benini:2014mia,Closset:2015rna} for a simple surface operator in 4d $SU(3)$ SCFT. When the 4d theory is super Yang-Mills, it is a duality in the Grassmannian theory \cite{Hori:2006dk}. We find a solution to the chiral ring equation of the dual $SU(2)$ theory and find an agreement with the $U(1)$ theory. From the instanton partition function, the dual twisted superpotential can be obtained by dual Higgsing prescription. In 3d/5d theories, the duality holds for a specific value of the Chern-Simons coefficient. In the example we are considering, it turns out to be zero, whereas in the case of 5d SYM, it was found to be $\pm1/2$. 

In section \ref{sec:chiral}, we discuss two dimensional sigma models coupled to 4d gauge theories. The solutions to the chiral ring equations are computed in small couplings and shown to be equivalent to the derivative with respect to the FI parameters of the twisted superpotential evaluated at the solution. In section \ref{sec:3d}, the 3d sigma model description of surface defects in 5d gauge theories are considered. Similar analysis is done with the 5d resolvents in place of the 4d resolvents. In section \ref{sec:inst}, the vacuum solutions of the previous sections are reproduced from the instanton partition functions. In section \ref{sec:duality}, evidences of the duality are given both from the perspective of chiral ring equations and from instanton partition functions. In section \ref{sec:dis}, we conclude with a few remarks for further study. In appendix \ref{sec:app}, some results of quantum vev's are listed.

\section{Chiral rings for 2d GLSM coupled to 4d $\cN=2$ theory}
\label{sec:chiral}

A surface operator as a monodromy defect is classified by a set of $M$ integers $n_i$ with $\sum_{i=1}^{M} n_i=N$. The surface operator denoted as $SU(N)[n_1,...,n_M]$ breaks the $SU(N)$ gauge group into a Levi subgroup $\mathbb{L}$,
\begin{equation}\label{levi}
\mathbb{L} = S[U(n_1) \times ... \times U(n_M)]
\end{equation}
on its world volume. The 2d description of the defect is an $\cN=(2,2)$ quiver GLSM with rank $r_I$ of node $I$ given by
\begin{equation}
r_I = \sum_{i=1}^{I} n_i \,.
\end{equation}
A bi-fundamental field connects each node and integrating them out leads to the effective twisted superpotential. 

In this section, we focus on the 2d quiver GLSMs, for which the sum of matter charges vanishes and coupled to a 4d theory. Simple surface operators with $\mathbb{L}=S[U(1) \times U(N-1)]$ in $\cN=2$ $SU(N)$ SCFT with $N_f=2N$ hypermultiplets are a class of defects described by these theories. For a quiver GLSM, we study a specific example that corresponds to a full surface operator with $\mathbb{L}=S[U(1)^3]$ in $SU(3)$ SQCD with $N_f=4$.

\subsection{Sigma models for simple surface operators in $\cN=2$ SCFT}
\label{sec:simplescft}

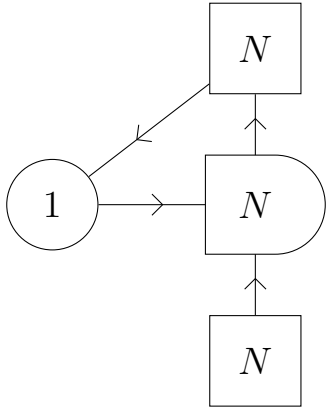
\begin{figure}[ht]
	\begin{center}
		\begin{tikzpicture}[decoration={markings, mark=at position 0.6 with {\draw (-4pt,-4pt) -- (0pt,0pt);       
				\draw (-4pt,4pt) -- (0pt,0pt);}}]
		\matrix[row sep=8mm,column sep=7mm] {
			&& \node(lflav)[flavor]{\Large $N$} ;\\
			\node(glast)[gauge] {\Large $1$};& &\node(gfN)[rotate=-90,gaugedflavor]{\rotatebox{90}{\Large $N$}};\\
			 && \node(rflav)[flavor]{\Large $N$} ;\\
		};
		\graph{(glast)--[postaction={decorate}](gfN); (rflav)--[postaction={decorate}](gfN); (gfN)--[postaction={decorate}](lflav)--[postaction={decorate}](glast)};
		\end{tikzpicture}
	\end{center}
	\vspace{-0.5cm}
	\caption{$U(1)$ quiver of a simple surface operator in $SU(N)$ gauge theory with $N_f=2N$.}
	\label{simplequiver}
\end{figure}

A simple surface operator in 4d $SU(N)$ SCFT can be described by the 2d GLSM with $N$ chirals of +1 charge with twisted mass $\Phi$ which is the scalar vev of 4d vector multiplet and $N$ chirals of -1 charge with twisted mass $m_f$ which is the mass of 4d hypermultiplet. Fig. \ref{simplequiver} shows the quiver diagram of the system.

The twisted superpotential for the sigma model is given by
\begin{equation}\label{eq:Ws}
W = 2\pi i \tau_1 \sigma + \sum_{f=1}^{N} \omega(\sigma - m_f) - \Big\langle \mathrm{Tr} \,\omega(\sigma - \Phi)\Big\rangle \,,
\end{equation}
where $\tau_1$ is the complexified FI parameter, Tr denotes the trace over the matrix $\Phi$, $\sigma$ is the scalar in the twisted field strength multiplet $\Sigma$, for which we have made no distinction in writing the superpotential for notational convenience. The function $w(x)$ is defined by
\begin{equation}
w(x) \equiv x\left(\log\frac{x}{\mu}-1\right) \,,
\end{equation}   
with an energy scale $\mu$. The brackets in the last term of $W$ denote the quantum vev of the chiral field $\Phi$ in the 4d gauge theory. Vacua of the theory are determined by \cite{Nekrasov:2009uh,Nekrasov:2009ui,Nekrasov:2009rc}
\begin{equation}
\exp\left(\frac{\p W}{\p \sigma}\right)=1 
\end{equation}
which results in the chiral ring equation
\begin{equation}\label{simvac}
q_1\prod_{f=1}^{N}(\sigma-m_f) = \mu^N\exp \left\langle \mathrm{Tr}\log\frac{\sigma-\Phi}{\mu} \right\rangle \,,
\end{equation}
where $q_1=e^{2\pi i \tau_1}$. The righthand side is the integrated resolvent of the four dimensional $SU(N)$ SCFT with $N_f=2N$ \cite{Gaiotto:2013sma},
\begin{equation}\label{simresol}
\exp\left\langle \mathrm{Tr}\log\frac{\sigma-\Phi}{\mu} \right\rangle = \frac{(1+q)P_N(\sigma)+\sqrt{(1+q)^2P_N^2(\sigma) - 4q B_{N_f}(\sigma)}}{2\mu^N} \,,
\end{equation}
where $P_N(\sigma)$ and $B_{N_f}(\sigma)$ take the form
\begin{eqnarray}
P_N(\sigma) &=& \prod_{i=1}^{N}(\sigma - e_i) \label{PN} \,,\\
B_{N_f}(\sigma) &=& B_L(x)B_R(x)\,, \label{BN}
\end{eqnarray}
with $q$ the dimensionless gauge coupling of the four dimensional SCFT (with mass deformation) and $e_i$ the quantum vev and
\begin{equation}
B_L(\sigma) = \prod_{f=1}^{N}(\sigma-m_f) \,,\qquad  B_R(\sigma) = \prod_{f=1}^{N}(\sigma+\tilde{m}_f)\,.
\end{equation}
Eq. (\ref{simvac}) and (\ref{simresol}) lead to the following twisted chiral ring equation
\begin{eqnarray}\label{simplechiral}
(1+q)P_N(\sigma) = q_1 B_L(\sigma) + \frac{q}{q_1} B_R(\sigma)\,.
\end{eqnarray} 

Note that the two-dimensional scale $\mu$ drops from the equation, due to the same number of chiral and antichiral. The twisted chiral ring relation \eqref{simplechiral} holds for all values of $m_f$ and so it can be set to zero. 

The $e_i$'s can be found from the Seiberg-Witten curve of the bulk theory \cite{Seiberg:1994rs, Seiberg:1994aj}, which can be written in the form 
\begin{equation}
y^2 = (1+q)^2P_N^2(x) - 4qB_{N_f}(x) \,,
\end{equation}
with the SW differential, 
\begin{equation}
\lambda = xd \log\left((1+q)P_N(x)+y\right) \,.
\end{equation}
Once $e_i$ is obtained, one can solve the chiral ring equation for $\sigma$ order by order in $q_1$ and $q$. 

Since the twisted superpotential $W_*(\sigma_*(q_1, q), q_1)$ evaluated on a solution $\sigma_*(q_1, q)$ satisfies 
\begin{equation}
\left.\frac{\p W}{\p \sigma}\right| _{\sigma_*} = 0 \,,
\end{equation} 
one obtains the following relation \cite{Ashok:2017lko}.
\begin{equation}\label{simplerel}
q_1\frac{d W_*}{d q_1} = q_1\frac{\p W_*}{\p q_1} = \sigma_*
\end{equation} 
This can be compared with the corresponding equation computed from ramified instanton partition function.

\subsubsection*{$U(1)$ sigma model coupled to $SU(2)$ SCFT}

A simple surface operator in $SU(2)$ SCFT with $N_f=4$ hypermultiplets corresponds to the two-dimensional $U(1)$ gauge theory with four matter fields of charge 1,1,-1,-1 coupled to the four-dimensional theory. The $SU(2)$ SCFT is coupled to the vacuum manifold of the 2d theory, which is the total space of $\cO(-1)\oplus\cO(-1)$ over $\mathbb{CP}^1$. The chiral ring equation \eqref{simplechiral} for the surface defect is written 
\begin{equation}\label{eq:simsu2}
(1+q)\prod_{i=1}^2 (\sigma-e_i) = q_1\prod_{f=1}^{2}(\sigma-m_f) + \frac{q}{q_1}\prod_{f=1}^{2}(\sigma+\tilde{m}_f) \,.
\end{equation} 
In Appendix \ref{qv4d}, $e_i$'s are computed in terms of $a$ and $q$. 
With a choice of classical vacuum, $a$, we write $\sigma$ as
\begin{equation}
\sigma = a + \sum_{i,j=1} s_{ij}q_1^i q^j \,,
\end{equation}
and solve for $s_{ij}$. Then one finds
\begin{eqnarray}\label{su2sol}
\sigma_* &=& a + \frac{q_1}{2a} \prod_{i=1}^2(a-m_i) + \frac{q}{2aq_1} \prod_{i=1}^2(a+\tilde{m}_i)  \nonumber\\
&&+ \frac{q_1^2}{8a^3}  (3a^2 +a(m_1+m_2) -m_1m_2)\prod_{i=1}^2(a-m_i)  \nonumber\\
&&+ \frac{q^2}{8a^3 q_1^2}  (3a^2 -a(\tilde{m}_1+\tilde{m}_2) -\tilde{m}_1\tilde{m}_2)\prod_{i=1}^2(a+\tilde{m}_i) + \cdots \,,
\end{eqnarray}
where $\sum m_f=\sum \tilde{m}_f=0$ can be imposed. 

 One can see from the solution that one instanton decomposes into a vortex and an instanton/anti-vortex contribution. At the second order, two instantons ramify into two vortices and two instantons/two anti-vortices \cite{Gorsky:2017hro}. 
When four dimensional dynamics is decoupled by $q \rightarrow 0$, the solution consists only of vortex contributions. 

Limits for theories of less flavors are well-defined in \eqref{eq:simsu2} and \eqref{su2sol}. For example, when all mass parameters become large such that  $q_1m_1m_2 \rightarrow \Lambda_1^2$, $\frac{q}{q_1}\tilde{m}_1\tilde{m}_2 \rightarrow \frac{\Lambda^4}{\Lambda_1^2}$ and $q \rightarrow 0$, where $\Lambda_1$ and $\Lambda$ are strong coupling scales for two and four dimensional theory, respectively as in \cite{Ashok:2017lko}, the solution reduces to the massive vacuum of $\mathbb{CP}^1$ sigma model coupled to pure super Yang-Mills. In this limit, the sigma model corresponds to $SU(2)[1,1]$ surface operator in super Yang-Mills.

\subsubsection*{$U(1)$ sigma model coupled to $SU(3)$ SCFT}
\label{sec:simplesu3}

In this case, the sigma model is a $U(1)$ theory with three chirals of +1 charge and three chirals of -1 charge and it is coupled to 4d $SU(3)$ gauge theory with three fundamental hypermultiplets and three anti-fundamental hypermultiplets. This is an extension of a Grassmanian sigma model for $SU(3)[1,2]$ surface operator coupled to SYM by including matter hypermultiplets. The chiral ring equation \eqref{simplechiral} are written
\begin{equation}
(1+q)\prod_{i=1}^3 (\sigma-e_i) = q_1\prod_{f=1}^{3}(\sigma-m_f) + \frac{q}{q_1}\prod_{f=1}^{3}(\sigma+\tilde{m}_f) \,.
\end{equation} 
For simplicity, we consider massless limit, $m_f \rightarrow 0$ and $\tilde{m}_f\rightarrow 0$.  
With $e_i$'s given in Appendix \ref{qv4d}, the massive vacuum for a choice of classical vacuum $a_1$ is found to be
\begin{eqnarray}\label{eq:simplesu3sol}
\sigma_* &=& a_1 + \frac{a_1^3}{(a_1-a_2)(a_1-a_3)}\left(q_1 + \frac{q}{q_1} \right)  \nonumber\\
&& + \frac{a_1^5 (a_1^2 + 3 a_2 a_3 - 2 a_1 (a_2 + a_3))}{(a_1 - a_2)^3 (a_1 - a_3)^3}\left( q_1^2+ \frac{q^2}{q_1^2}\right) + \cdots 
\end{eqnarray}

This 2d/4d system has dual description, where the 2d gauge group is $SU(2)$, as shown in Sec. \ref{sec:duality}.

\subsection{Sigma model for full surface operator in SQCD}
\label{sec:quiver}

\begin{figure}
	\begin{center}
		\begin{tikzpicture}[decoration={markings, mark=at position 0.6 with {\draw (-4pt,-4pt) -- (0pt,0pt);       
				\draw (-4pt,4pt) -- (0pt,0pt);}}]
		\matrix[row sep=9mm,column sep=6mm] {
			 && && \node(lflav)[flavor]{\Large $2$} ;\\
			\node (g1)[gauge] {\Large $1$};  && \node(g2)[gauge] {\Large $2$};  && \node(gfN)[rotate=-90,gaugedflavor]{\rotatebox{90}{\Large $3$}};\\
			 && && \node(rflav)[flavor]{\Large $2$} ;\\
		};
		\graph{(g1) --[postaction={decorate}](g2)--[postaction={decorate}](gfN); 
			(rflav)--[postaction={decorate}](gfN); 
			(gfN)--[postaction={decorate}](lflav)--[postaction={decorate}](g1);(lflav)--[postaction={decorate}](g2)
			(rflav)--[postaction={decorate}](g2);(rflav)--[postaction={decorate}](g1)
		};
		\end{tikzpicture}
	\end{center}
	\vspace{-0.5cm}
	\caption{Quiver diagram of a full surface operator in SU$(3)$ SQCD with 4 flavors. The arrow from flavor group to each 2d gauge node represents one chiral multiplet. }
	\label{quiversu3}
\end{figure}
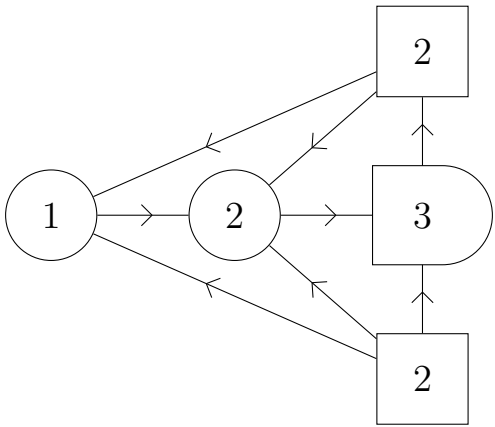

As a quiver sigma model example, we focus on a full surface operator in $SU(3)$ theory with four flavors. The 2d quiver theory has $U(1)\times SU(2)$ gauge group and the matter contents consist of a bifundamental between the two nodes, two chirals of +1 charge at each node from the bulk hypermultiplets in symmetric way, and at the $SU(2)$ node three chirals of -1 charge coupled to the four dimensional $SU(3)$ gauge group. Fig.~\ref{quiversu3} shows the quiver diagram. When the masses of hypermultiplets become large, this system approaches to the sigma model of $SU(3)[1,1,1]$ surface operator, whose target space is a flag variety, coupled to super Yang-Mills.

The twisted superpotential is given by
\begin{eqnarray}\label{quiversu3W}
W &=& 2\pi i\tau_1\sigma + 2\pi i\tau_2\sum_{i=1}^{2}\tilde{\sigma}_i - \sum_{i=1}^{2}w(\sigma-\tilde{\sigma}_i) + w(\sigma-m_1) + w(\sigma+\tilde{m}_1) \nonumber\\
&&  + \sum_{i=1}^{2}\Big(w(\tilde{\sigma}_i-m_2) + w(\tilde{\sigma}_i+\tilde{m}_2)\Big) - \sum_{i=1}^{2}\Big\langle\mathrm{Tr}\,w(\tilde{\sigma}_i-\Phi)\Big\rangle \,,
\end{eqnarray}
where $\tau_I$ is the FI parameter of $I$-th node and $\tilde{\sigma}_i$ the diagonal entry of the scalar in the twisted chiral multiplet of the $SU(2)$ node. From 
\begin{equation}
\exp\left(\frac{\p W}{\p\sigma}\right)= \exp\left(\frac{\p W}{\p\tilde{\sigma}_i}\right)=1 \,,
\end{equation}
the twisted chiral ring relations are
\begin{eqnarray}\label{eq:quiverchi1}
\prod_{i=1}^{2}(\sigma-\tilde{\sigma}_i) &=& q_1(\sigma-m_1)(\sigma + \tilde{m}_1) \,,\\
\exp\Big\langle\mathrm{Tr}\,\log(\tilde{\sigma}_i-\Phi)\Big\rangle &=& q_2(\sigma-\tilde{\sigma}_i)(\tilde{\sigma}_i - m_2)(\tilde{\sigma}_i+\tilde{m}_2) \,,
\label{eq:quiversu3}
\end{eqnarray}
where $q_I = e^{2\pi i\tau_I}$. The four-dimensional theory couples not just to the $SU(2)$ node but to the first node as well.  

When the four-dimensional theory is decoupled, the second equation simplifies to 
\begin{equation}\label{flag2}
\prod_{j=1}^{3}(\tilde{\sigma}_i-\Phi_j) = q_2(\sigma-\tilde{\sigma}_i)(\tilde{\sigma}_i - m_2)(\tilde{\sigma}_i+\tilde{m}_2) \,,
\end{equation}
where $\Phi_i$ is diagonal entry of $\Phi$. In the limit of large masses such that $q_1m_1\tilde{m}_1\rightarrow -\Lambda_1^2$, $q_2m_2\tilde{m}_2\rightarrow -\Lambda_2^2$, where $\Lambda_I$ is the UV scale of node $I$, the chiral ring relations \eqref{eq:quiverchi1} and \eqref{flag2} reduce to the chiral rings of the sigma model for a flag variety. 

From the four-dimensional resolvent, one has
\begin{equation}\label{resol3}
\exp\Big\langle\mathrm{Tr}\,\log(\tilde{\sigma}_i-\Phi)\Big\rangle = \frac{P_3(\tilde{\sigma}_i)+\sqrt{P_3^2(\tilde{\sigma}_i) - 4\Lambda^2B_4(\tilde{\sigma}_i)}}{2} \,,
\end{equation}
where $\Lambda$ is the strong coupling scale of the 4d theory.
The quantum vev $e_i$ can be computed from the SW curve 
\begin{equation}
y^2 = P_3^2(x) - 4\Lambda^2B_4(x) \,,
\end{equation}
and they are listed in Appendix \ref{qv4d}. 
Using \eqref{resol3}, the second equation of the chiral ring relations \eqref{eq:quiversu3} can be written as
\begin{equation}\label{eq:quiverchi2}
P_3(\tilde{\sigma}_i) = q_2(\sigma-\tilde{\sigma}_i)(\tilde{\sigma}_i-m_2)(\tilde{\sigma}_i+\tilde{m}_2) + q_1 q_3\frac{(\tilde{\sigma}_i-m_1)(\tilde{\sigma}_i+\tilde{m}_1)}{(\sigma-\tilde{\sigma}_i)} \,,
\end{equation} 
where we have defined $\Lambda^2 \equiv q_1q_2q_3$. 

A massive vacuum can be found with a natural choice of classical vacua by writing 
\begin{eqnarray*}
\sigma &=& a_1 + \sum_{i,j,k} s_{ijk}\,q_1^i q_2^j q_3^k \,,\nonumber\\
\tilde{\sigma}_i &=& a_i + \sum_{i,j,k} \tilde{s}_{ijk}\,q_1^iq_2^jq_3^k \,,
\end{eqnarray*}
for $i=1,2$. Solutions to \eqref{eq:quiverchi1} and \eqref{eq:quiverchi2} are found 
\begin{eqnarray}\label{quiversu3sig}
\sigma_* &=& a_1 + \frac{(a_1-m_1)(a_1+\tilde{m}_1)}{a_1 - a_2}q_1 - \frac{1}{a_1-a_3}q_3 \nonumber\\
&+& \frac{(a_1-m_1)(a_1+\tilde{m}_1)(a_1^2-2a_1a_2 +a_2(m_1-\tilde{m}_1)+m_1\tilde{m}_1)}{(a_1 - a_2)^3}q_1^2 - \frac{1}{(a_1-a_3)^3}q_3^2 \nonumber\\
&-&  \frac{(a_1-m_1)(a_1+\tilde{m}_1)(a_1(a_2-a_3) -a_2a_3 +a_3(m_2-\tilde{m}_2)+ m_2\tilde{m}_2)}{(a_1 - a_2) (a_1 - a_3) (a_2 - a_3)}q_1q_2  \nonumber\\
&+& \frac{(a_2-m_2)(a_2+\tilde{m}_2)}{(a_1 - a_2) (a_1 - a_3) (a_2 - a_3)}q_2q_3 + \cdots 
\end{eqnarray}
and
\begin{eqnarray}\label{quiversu3sig2}
\sum_{i=1}^{2}\tilde{\sigma}_{*i} &=& a_1+a_2 +\frac{(a_2-m_2)(a_2+\tilde{m}_2)}{a_2-a_3}q_2-\frac{1}{a_1-a_3}q_3 \nonumber\\
&+& \frac{(a_2-m_2)(a_2+\tilde{m}_2)(a_2^2-2a_2a_3 +a_3(m_2-\tilde{m}_2) +m_2\tilde{m}_2))}{\left(a_2-a_3\right)^3}q_2^2 -\frac{1}{\left(a_1-a_3\right)^3}q_3^2  \nonumber\\
&-& \frac{(a_1-m_1)(a_1+\tilde{m}_1)(a_1(a_2-a_3)-a_2a_3 +a_3(m_2-\tilde{m}_2) +m_2\tilde{m}_2)}{\left(a_1-a_2\right) \left(a_1-a_3\right) \left(a_2-a_3\right)}q_1q_2 \nonumber\\
&+& \frac{a_1 a_2-a_1a_3+a_2 a_3 -a_2(m_1-\tilde{m}_1)-m_1\tilde{m}_1}{\left(a_1-a_2\right) \left(a_1-a_3\right) \left(a_2-a_3\right)}q_1q_3 + \cdots
\end{eqnarray} 
As mentioned in \eqref{simplerel}, the on-shell twisted superpotential \eqref{quiversu3W} satisfies
\begin{eqnarray}
q_1\frac{d W_*}{d q_1} &=& \sigma_* \\
q_2\frac{d W_*}{d q_2} &=& \sum_{i=1}^{2}\tilde{\sigma}_{*i},.
\end{eqnarray}
These relations will be used to compare the results in Sec. \ref{sec:inst}.

\section{3d sigma models coupled to 5d $\cN=1$ theories}
\label{sec:3d}

The theories considered in the previous section can be lifted to three-dimensional quivers coupled to five-dimensional $\cN=1$ theories compactified on a circle. A Kaluza-Klein mode on a circle of circumference $\beta$ contributes to the twisted superpotential by $w(x+2\pi n/\beta)=(x+2\pi n/\beta)\log((x+2\pi n/\beta)/\mu-1)$ for an integer $n$. The regularized sum can be defined
\begin{eqnarray}
L(x) &\equiv& \sum_{n \in \mathbb{Z}} w\left(x+\frac{2\pi in}{\beta}\right) 
\end{eqnarray}	
such that 
\begin{equation}
\frac{\p L(x)}{\p x} = \sum_{n\in\mathbb{Z}}\log\left(\frac{x}{\mu}+\frac{2\pi in}{\beta\mu}\right) = \log\left(2\sinh\frac{\beta x}{2}\right)
\end{equation}
and as $\beta \rightarrow 0$,
\begin{equation}
L(x) \rightarrow x(\log \beta x -1) \,.
\end{equation}
Since $\mu$ is the UV scale, the identification $\beta=1/\mu$ can be made and $L(x)$ goes back to $w(x)$ in the $\beta \rightarrow 0$ limit \cite{Ashok:2017lko}. The function $L(x)$ plays the role of $w(x)$ in the previous section, so the twisted superpotential is written in terms of $L(x)$ in place of $w(x)$. 

In 3 and 5 dimension, Chern-Simons terms can be present in the twisted superpotential \cite{Ashok:2017bld}. In the following, we will set 5d Chern-Simons terms to zero and discuss 3d Chern-Simons terms only.

\subsection{Sigma models for simple defects in 5d $SU(N)$ theory with $N_f=2N$}

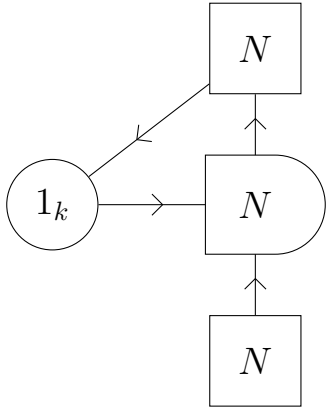
\begin{figure}[ht]
	\begin{center}
		\begin{tikzpicture}[decoration={markings, mark=at position 0.6 with {\draw (-4pt,-4pt) -- (0pt,0pt);       
				\draw (-4pt,4pt) -- (0pt,0pt);}}]
		\matrix[row sep=8mm,column sep=7mm] {
			&& \node(lflav)[flavor]{\Large $N$} ;\\
			\node(glast)[gauge] {\Large $1_k$};& &\node(gfN)[rotate=-90,gaugedflavor]{\rotatebox[]{90}{\Large $N$}};\\
			&& \node(rflav)[flavor]{\Large $N$} ;\\
		};
		\graph{(glast)--[postaction={decorate}](gfN); (rflav)--[postaction={decorate}](gfN); (gfN)--[postaction={decorate}](lflav)--[postaction={decorate}](glast)};
		\end{tikzpicture}
	\end{center}
	\vspace{-0.5cm}
	\caption{$U(1)$ quiver of a simple defect in $SU(N)$ gauge theory with $N_f=2N$. The subscript $k$ denotes the Chern-Simons coefficient. }
	\label{3dsimplequiver}
\end{figure}

A simple defect in 5d $\cN=1$ $SU(N)$ theory with $2N$ matter hypermultiplets has a description of the 3d lift of the sigma model for the simple surface operator in Sec. \ref{sec:simplescft}. The quiver diagram is illustrated in Fig.~\ref{3dsimplequiver}. 

One can write the twisted superpotential as \cite{Nekrasov:2009uh,Chen:2012we}
\begin{equation}\label{eq:3dsimpleW}
W = 2\pi i\tau_1\sigma -\frac{\beta k}{2}\sigma^2 + \sum_{f=1}^{N}L(\sigma-m_f) - \Big\langle \mathrm{Tr}L(\sigma-\Phi) \Big\rangle\,,
\end{equation}
where $\Phi = A_0+i\phi$ in terms of the 5d gauge field in the circle direction and 5d real scalar in the $\cN=1$ vector multiplet. The second term is the contribution from the Chern-Simons term, coming from integrating out massive chiral fields.  
The chiral ring relation becomes 
\begin{equation}\label{eq:5dsimple}
q_1e^{-\beta k\sigma}\prod_{f=1}^{N}2\sinh\frac{\beta(\sigma-m_f)}{2} = \exp\Big\langle\mathrm{Tr}\log\left(2\sinh\frac{\beta(\sigma-\Phi)}{2}\right)\Big\rangle \,.
\end{equation}

The effect of coupling to five dimensional theory can be seen through the five dimensional SW curve \cite{Nekrasov:1996cz,Nekrasov:2002qd,Nekrasov:2003rj},
\begin{equation}
Y^2 = (1+q)^2\widehat{P}_N^2(x) - 4q\widehat{B}_{N_f}(x) \,,
\end{equation}
where the polynomials $\widehat{P}_N(x)$ and $\widehat{B}_{N_f}(x)$ are
\begin{eqnarray}\label{5dPN}
\widehat{P}_N(x) &=& \prod_{i=1}^{N}2\sinh\frac{\beta (x-\widehat{e}_i)}{2}\,,\nonumber\\
\widehat{B}_{N_f}(x) &=&  \prod_{f=1}^{N}\left(2\sinh\frac{\beta(x-m_f)}{2}\right)  \prod_{f=1}^{N}\left(2\sinh\frac{\beta(x+\tilde{m}_f)}{2}\right)  \,.
\end{eqnarray}
The quantum vev $\widehat{e}_i$ is obtained by integrating the SW differential, 
\begin{equation}\label{5dsw1}
\widehat{\lambda} = x\,d\log((1+q)\widehat{P}_N(x)+Y) \,,
\end{equation}
which is done in Appendix \ref{qv5d}.
The five dimensional resolvent is defined by \cite{Wijnholt:2004rg}
\begin{eqnarray}
\widehat{T}(x) &\equiv& \left\langle \mathrm{Tr}\coth\frac{\beta(x-\Phi)}{2}\right\rangle \nonumber\\
&=& \frac{2}{\beta}\frac{\p}{\p x}\left\langle\mathrm{Tr}\log\left(2\sinh\frac{\beta(x-\Phi)}{2}\right)\right\rangle \,.
\end{eqnarray}
From the SW differential \eqref{5dsw1}, the resolvent can also be written as
\begin{equation*}
\widehat{T}(x) = \frac{2}{\beta}\left(\frac{(1+q)\widehat{P}_N'}{\sqrt{(1+q)^2\widehat{P}_N^2-4q\widehat{B}_{N_f}}} - \frac{(1+q)\widehat{P}_N\widehat{B}_{N_f}'}{2\widehat{B}_{N_f}\sqrt{(1+q)^2\widehat{P}_N^2-4q\widehat{B}_{N_f}}} +\frac{\widehat{B}'_{N_f}}{2\widehat{B}_{N_f}}\right) \,,
\end{equation*}
where $'$ denotes the derivative with respect to $x$.
Thus, the right-hand side of \eqref{eq:5dsimple} satisfies
\begin{equation*}
\exp\left\langle \mathrm{Tr}\log \left(2\sinh\frac{\beta(\sigma-\Phi)}{2}\right) \right\rangle = \frac{(1+q)\widehat{P}_N(\sigma)+\sqrt{(1+q)^2\widehat{P}_N^2(\sigma) - 4q \widehat{B}_{N_f}(\sigma)}}{2} \,,
\end{equation*}  
and the chiral ring equation for the defect is 
\begin{equation}\label{5dsimple}
(1+q)\widehat{P}_N(\sigma) = q_1S^{-k}\widehat{B}_L(\sigma) + \frac{q}{q_1}S^k\widehat{B}_R(\sigma) \,,
\end{equation}
where $S\equiv e^{\beta\sigma}$ and
\begin{equation}
\widehat{B}_L(x)=\prod_{f=1}^{N}2\sinh\frac{\beta(x-m_f)}{2} \quad\,,\quad \widehat{B}_R(x)=\prod_{f=1}^{N}2\sinh\frac{\beta(x+\tilde{m}_f)}{2} \,.
\end{equation}
so that $\widehat{B}_{N_f}=\widehat{B}_L\widehat{B}_R$.
Note that the four dimensional limit $\beta \rightarrow 0$ of \eqref{5dsimple} is well defined and consistent with the chiral ring in 4d, \eqref{simplechiral}. 

The limit of SYM can be taken by letting $|m_f|$ and $|\tilde{m}_f|$ to infinity with a suitable choice of Chern-Simons coefficient. 
Since the mass parameters satisfy 
\begin{equation}
\sum_{f=1}^{N}m_f= \sum_{f=1}^{N}\tilde{m}_f=0 \,,
\end{equation}
one can take $p$ of the $m_f$'s to $+\infty$ and $N-p$ of them to $-\infty$. Similarly for $\tilde{m}_f$'s. Then $\widehat{B}_L(\sigma)$ becomes proportional to the factor of $e^{\frac{\beta}{2}(2p-N)\sigma}$ and $\widehat{B}_R(\sigma)$ to $e^{-\frac{\beta}{2}(2p-N)\sigma}$. For the SYM limit, these $\sigma$ dependent factors must be canceled. Thus, Chern-Simons factors for
\begin{equation}\label{cs}
k=\frac{2p-N}{2} \,,
\end{equation}
should be chosen for the sigma model of a simple defect when coupled to SYM. For example, $k=0$ for $N=2$ and $k=\pm\frac{1}{2}$ for $N=3$.

The logarithmic derivative of the twisted superpotential \eqref{eq:3dsimpleW} with respect to the exponential of FI parameter evaluated at a vacuum solution $\sigma_*$ has the following property
\begin{equation}
q_1\frac{d W_*}{dq_1} = \sigma_*
\end{equation}
as before.  

\subsubsection*{$U(1)$ sigma model coupled to 5d $SU(2)$ theory with $N_f=4$}

We consider the sigma model with flavor symmetry $SU(2)\times SU(2)$ for the simple defect in five dimension. The matter charges are 1,1,-1,-1 as in the 2d/4d example. The chiral ring relation is given by \eqref{5dsimple} with $N=2$ and $k=0$. A vacuum solution is found by using the expression of $\widehat{e}_i$ given in Appendix \ref{qv5d}
\begin{eqnarray}\label{5dsu2sol}
\sigma_* &=& a + \frac{\widehat{B}_L(a)}{2\beta\sinh\beta a}q_1 +  \frac{\widehat{B}_R(a)}{2\beta\sinh\beta a}\frac{q}{q_1} \nonumber\\
&+& \frac{\widehat{B}_L(a)\left(\sinh\frac{\beta(a+m_1)}{2}\sinh\frac{\beta(a-m_2)}{2}+\sinh\beta a\sinh\frac{\beta(a-m_1)}{2}\cosh\frac{\beta(a-m_2)}{2}\right)}{2\beta\sinh^3\beta a}q_1^2 \nonumber\\
&+& \frac{\widehat{B}_R(a)\left(\sinh\frac{\beta(a-\tilde{m}_1)}{2}\sinh\frac{\beta(a+\tilde{m}_2)}{2}+\sinh\beta a\sinh\frac{\beta(a+\tilde{m}_1)}{2}\cosh\frac{\beta(a+\tilde{m}_2)}{2}\right)}{2\beta\sinh^3\beta a}\frac{q^2}{q_1^2} \nonumber\\
&+&  \cdots
\end{eqnarray}
As $\beta\rightarrow 0$, we recover the 2d result \eqref{su2sol}.

To see the SYM limit, note that the mass parameters satisfy $m_1+m_2=\tilde{m}_1+\tilde{m}_2=0$. So for the limit of  $m_1,\tilde{m}_2\rightarrow +\infty$, the dominant factors $e^{\beta(m_1-m_2)/2}q_1$ and $e^{-\beta(\tilde{m}_1-\tilde{m}_2)/2}(q/q_1)$ are taken to $-\beta^2\Lambda_1^2$ and $-\beta^2\Lambda^4/\Lambda_1^2$, respectively. The above $\sigma_*$ then becomes 
\begin{equation}
\sigma_* = a + \frac{\beta}{2\sinh\beta a}\left(\Lambda_1^2+\frac{\Lambda^4}{\Lambda_1^2}\right) 
-  \frac{\beta^3\cosh\beta a}{8\sinh^3\beta a}\left(\Lambda_1^4+\frac{\Lambda^8}{\Lambda_1^4}\right) + \cdots \,,
\end{equation}
which is the vacuum solution for $SU(2)[1,1]$ defect in 5d super Yang-Mills given in \cite{Ashok:2017lko}.

\subsection*{$U(1)$ sigma model coupled to 5d $SU(3)$ theory with $N_f=6$}

The flavor symmetry of the sigma model is $SU(3)\times SU(3)$, where the matter contents are three fundamental chirals and three anti-fundamental chirals. We take massless limit again with vanishing Chern-Simons term for simplicity. A solution to the chiral ring equation is obtained 
\begin{eqnarray}\label{5dsu3sol}
\sigma_* &=& a_1 + \frac{2\sinh^3\frac{\beta a_1}{2}}{\beta \sinh\frac{\beta(a_1-a_2)}{2}\sinh\frac{\beta(a_1-a_3)}{2}}\left(q_1+\frac{q}{q_1}\right) \nonumber\\
&+& \frac{\sinh^5\frac{\beta a_1}{2}}{32\beta\sinh^3\frac{\beta(a_1-a_2)}{2}\sinh^3\frac{\beta(a_1-a_3)}{2}}\left(5\cosh\frac{\beta(a_1-a_2-a_3)}{2}+\cosh\frac{\beta(3a_1-a_2-a_3)}{2}\right.\nonumber\\
&& \hskip2cm \left.-3\cosh\frac{\beta(a_1+a_2-a_3)}{2}-3\cosh\frac{\beta(a_1-a_2+a_3)}{2}\right)\left(q_1^2+\frac{q^2}{q_1^2}\right) \nonumber\\
&+& \cdots
\end{eqnarray}
The solution is consistent with its four-dimensional version \eqref{eq:simplesu3sol} in the limit of small $\beta$.

\subsection{Sigma model for full defect in 5d SQCD }

This is the case of $U(1)\times SU(2)$ sigma model coupled to 5d $SU(3)$ gauge theory with four flavors. The quiver diagram is drawn in Fig.\,\ref{quiversu3}. With zero Chern-Simons coefficient, the twisted superpotential is just \eqref{quiversu3W} with $w(x)$ replaced by $L(x)$,
\begin{eqnarray}\label{5dquiversu3W}
W &=& 2\pi i\tau_1\sigma + 2\pi i\tau_2\sum_{i=1}^{2}\tilde{\sigma}_i - \sum_{i=1}^{2}L(\sigma-\tilde{\sigma}_i) + L(\sigma-m_1) + L(\sigma + \tilde{m}_1) \nonumber\\
&&  + \sum_{i=1}^{2} \Big( L(\tilde{\sigma}_i-m_2) + L(\tilde{\sigma}_i+\tilde{m}_2) \Big) - \sum_{i=1}^{2}\Big\langle\mathrm{Tr}\,L(\tilde{\sigma}_i-\Phi)\Big\rangle \,.
\end{eqnarray}
Twisted chiral ring relations are in the form of \eqref{eq:quiverchi1} and \eqref{eq:quiverchi2} with factors of the hyperbolic sine function. For the $U(1)$ node, it is given by
\begin{equation}
\prod_{i=1}^{2}\left(2\sinh\frac{\beta(\sigma-\tilde{\sigma}_i)}{2}\right) = q_1\left( 2\sinh\frac{\beta(\sigma-m_1)}{2}\right) \left(2\sinh\frac{\beta(\sigma+\tilde{m}_1)}{2}\right) \,,
\end{equation}
For the $SU(2)$ node, we need data from the bulk theory. The SW curve is 
\begin{equation}
Y^2 = \widehat{P}_3^2(x) - 4(\beta\Lambda)^2\widehat{B}_4(x) \,,
\end{equation}
where $\Lambda$ is the strong coupling scale, and the characteristic polynomial $\widehat{B}_4(x)$ is given by
\begin{eqnarray}
\widehat{B}_4(x) &=& \prod_{f=1}^{2}\left(2\sinh\frac{\beta(\tilde{\sigma}_i-m_f)}{2}\right)\prod_{f=1}^{2}\left(2\sinh\frac{\beta(\tilde{\sigma}_i+\tilde{m}_f)}{2}\right) \,.
\end{eqnarray}
Using the relation from the 5d resolvent
\begin{equation}
\exp\left\langle \mathrm{Tr}\log \left(2\sinh\frac{\beta(\tilde{\sigma}_i-\Phi)}{2}\right) \right\rangle = \frac{\widehat{P}_3(\tilde{\sigma}_i)+\sqrt{\widehat{P}_3^2(\tilde{\sigma}_i) - 4(\beta\Lambda)^2 \widehat{B}_{4}(\tilde{\sigma}_i)}}{2} \,,
\end{equation} 
the chiral ring equation for the second node is written 
\begin{eqnarray}
\widehat{P}_3(\tilde{\sigma}_i) &=& q_2 \left(2\sinh\frac{\beta(\sigma-\tilde{\sigma}_i)}{2}\right)\left(2\sinh\frac{\beta(\sigma-m_2)}{2}\right)\left(2\sinh\frac{\beta(\sigma+\tilde{m}_2)}{2}\right) \nonumber\\
&+& q_1 q_3 \left(2\sinh\frac{\beta(\sigma-\tilde{\sigma}_i)}{2}\right)^{-1}\left(2\sinh\frac{\beta(\sigma-m_1)}{2}\right)\left(2\sinh\frac{\beta(\sigma+\tilde{m}_1)}{2}\right) \,,
\end{eqnarray}
with $q_3\equiv(\beta\Lambda)^2/q_1q_2$.

A vacuum solution is found to be
\begin{eqnarray}\label{eq:5dsu3quiversol}
\sigma_* &=& a_1 + \frac{2\sinh\frac{\beta(a_1-m_1)}{2}\sinh\frac{\beta(a_1+\tilde{m}_1)}{2}}{\beta\sinh\frac{\beta(a_1-a_2)}{2}}q_1 - \frac{1}{2\beta\sinh\frac{\beta(a_1-a_3)}{2}}q_3 +\cdots\,, \nonumber\\
\sum_{i=1}^{2}\tilde{\sigma}_{*i} &=& a_1+a_2 + \frac{2\sinh\frac{\beta(a_2-m_2)}{2}\sinh\frac{\beta(a_2+\tilde{m}_2)}{2}}{\beta\sinh\frac{\beta(a_2-a_3)}{2}}q_2 - \frac{1}{2\beta\sinh\frac{\beta(a_2-a_3)}{2}}q_3 +\cdots\,.\nonumber\\
\end{eqnarray}
When $\beta$ goes to zero, it reproduces the 4d result of \eqref{quiversu3sig} and \eqref{quiversu3sig2} for massless case.

\section{Twisted superpotential from instanton partition function}
\label{sec:inst} 

In this section, the twisted superpotential of the theories considered in the previous sections are derived from instanton partition function. Higgsing method is used for simple surface operators and orbifold construction for full surface operators. Then similar computations are carried out for the corresponding 3d/5d coupled systems. 

\subsection{Higgsing the instanton partition function}

In brane construction, $SU(N)\times SU(N)$ quiver theory with $N_f=2N$ can be engineered by a stack of $N$ D4 branes with three NS5 branes intersecting them. A simple surface operator can be realized by moving the middle NS5 brane in the transverse direction with a D2 brane stretched between the NS5 brane and one of the D4 branes. 

We introduce the notation 
\begin{equation}
E(x,Y,W, s) = x+\eone(w_{i}-j)-\etwo(y^T_{j}-i+1),
\end{equation}
where $s=(i,j)\in Y$ is a box in the Young diagram $Y$ with vertical position $i$ and horizontal position $j$. The quantity  $w_{i}$ is the length of $i$th row in $W$ and $y^T_{j}$ is the height of $j$th column in $Y$. 

The contribution of a bifundamental multiplet of mass $\mu$ charged under two gauge groups whose scalar vevs and Young diagrams are $\vec{a}$, $Y_m$ and $\vec{b}$, $W_n$ is given by
\begin{equation}
(E(a_m-b_n,Y_m,W_n,s)-\mu)(E(b_n-a_m,W_n,Y_m,t)+\mu+\epsilon) \,,
\end{equation}
with $\epsilon \equiv \eone+\etwo$. A matter hypermultiplet of mass $m_f$ in the fundamental representation contributes by a factor
\begin{equation}
F(a_m,Y_m,s,m_f) = a_m -(j-1)\eone - (i-1)\etwo + m_f \,.
\end{equation} 

The instanton partition function of $SU(N)\times SU(N)$ quiver theory with a bifundamental and $N$ fundamentals at each end can be written as \cite{Kozcaz:2010af}
\begin{eqnarray}\label{nekhiggs}
Z(\vec{a},\vec{b}) &=& \sum_{\vec{Y},\vec{W}}z^{|\vec{Y}|}\tilde{z}^{|\vec{W}|}\prod_{m,n=1}^{N}\prod_{s\in Y_m} \frac{(E(a_m-b_n,Y_m,W_n,s)-\mu)\prod_{f=1}^{N} F(a_m,Y_m,s,m_f)}{E(a_m-b_n,Y_m,Y_n,s)(E(a_m-b_n,Y_m,Y_n,s)+\epsilon)} \nonumber\\
&& \times \prod_{t\in W_m} \frac{(E(b_m-a_n, W_m,Y_n,t)+\mu+\epsilon)\prod_{f=1}^{N} F(b_m,W_m,t,\tilde{m}_f)}{E(b_m-b_n,W_m,W_n,t)(E(b_m-b_n,W_m,W_n,t)+\epsilon)} \,,
\end{eqnarray}  
where for the first gauge group, $z$ is the instanton counting parameter, $\vec{Y}=(Y_1,Y_2,...,Y_N)$ the $N$-tuple of Young diagrams, $|\vec{Y}|$ the total number of boxes in $\vec{Y}$, and $\vec{a}=(a_1,a_2,...,a_N)$ the vev of adjoint scalar. Similarly $\tilde{z}$, $\vec{W}$, $\vec{b}$, $|\vec{W}|$ are defined for the second gauge group. For $SU(N)$, the vevs satisfy $\sum_m a_m = \sum_n b_n = 0$. We also require the mass parameters be subject to $\sum m_f=\sum \tilde{m}_f=0$.  

The instanton partition function in the presence of a simple surface operator is obtained by setting the $b_n$ and $\mu$ to special values \cite{Dorey:2011pa},
\begin{equation}\label{higgscond}
\vec{b}=\vec{a}+(\etwo,0,0,...,0) \,, \qquad \mu=0 \,.
\end{equation} 
An equivalent prescription in the context of the AGT relation \cite{Alday:2009aq} is 
\begin{equation}
\vec{b} = \vec{a} + \etwo \kappa_1 \,, \qquad \mu = \frac{\etwo}{N} \,,
\end{equation}
where $\kappa_1 =\frac{1}{N}(N-1,-1,...,-1)$ is the fundamental weight of the $A_{N-1}$ Lie algebra \cite{Wyllard:2010vi}. 

After imposing these constraints, the twisted superpotential can be extracted from the instanton partition function for small $\Omega$ deformation parameters $\eone$ and $\etwo$
\begin{equation}
\log Z(\vec{a}) = -\frac{\cF}{\eone\etwo} + \frac{\cW}{\eone} + \cdots \,,
\end{equation}
where $\cF$ is the prepotential of the bulk theory and $\cW$ is the effective twisted superpotential of the degrees of freedom localized on the surface.

\subsection*{SU(2)[1,1] surface operator in $SU(2)$ SCFT}

The Higgsing prescription becomes
\begin{equation}
\vec{b} = \vec{a} + \etwo(1,0)\,, \quad \mu = 0 \,.
\end{equation}
The ramified instanton partition function is given by
\begin{equation}\label{neksu2}
Z[1,1] = 1 +\frac{(a-m_1)(a-m_2)}{\eone(2a+\eone+\etwo)}z + \frac{(a+\tilde{m}_1+\etwo)(a+\tilde{m}_2+\etwo)}{\eone(-2a +\eone)}\tilde{z} + \cdots\,,
\end{equation} 
where $a=a_1$. 
The parameters $z$ and $\tilde{z}$ are mapped to the sigma model parameters as
\begin{equation}\label{simpleparam}
z= q_1\,, \quad \tilde{z} = \frac{q}{q_1} \,.
\end{equation} 
The logarithmic derivative of $\cW$ with respect to $q_1$ yields 
\begin{eqnarray}
q_1\frac{d\cW}{d q_1} &=& \frac{1}{2a}(a-m_1)(a-m_2)q_1+\frac{1}{2a}(a+\tilde{m}_1)(a+\tilde{m}_2)\frac{q}{q_1} \nonumber\\
&& + \frac{(a-m_1)(a-m_2)(3a^2-m_1m_2)}{8a^3}q_1^2 +\frac{(a+\tilde{m}_1)(a+\tilde{m}_2)(3a^2-\tilde{m}_1\tilde{m}_2)}{8a^3}\frac{q^2}{q_1^2} \nonumber\\
&& +  \cdots\,,
\end{eqnarray}
which agrees with the result in \eqref{su2sol}.

\subsection*{SU(3)[1,2] surface operator in $SU(3)$ SCFT}

The Higgsing prescription is
\begin{equation}\label{higgssu3}
\vec{b} = \vec{a} + \etwo\left(1, 0, 0\right)\,, \qquad \mu = 0\,.
\end{equation}
The Higgsed instanton partition function is given by
\begin{eqnarray}\label{neksu3}
Z[1,2] &=& 1 + z\left( \frac{(\eone+2\etwo)\prod_{f=1}^3(a_1-m_f)}{\eone\etwo (a_1-a_2)(a_1-a_3)} +  \frac{\epsilon(a_1-a_2+\eone+2\etwo)\prod_{f=1}^{3}(a_2-m_f)}{\eone\etwo(a_2-a_1)(a_2-a_3)(a_1-a_2+\epsilon)}\right. \nonumber\\
&&  + \left.\frac{\epsilon(a_1-a_3+\eone+2\etwo)\prod_{f=1}^{3}(a_3-m_f)}{\eone\etwo(a_3-a_1)(a_3-a_2)(a_1-a_3+\epsilon)} \right) \nonumber\\
&& - \tilde{z}\frac{\prod_{f=1}^3(a_1+\tilde{m}_f+\etwo)}{\eone(a_2-a_1+\eone)(a_3-a_1+\eone)} +\cdots,
\end{eqnarray}
from which we obtain the effective twisted superpotential and its derivative. With the identification \eqref{simpleparam}, 
\begin{equation}
q_1\frac{d\cW}{dq_1} = \frac{\prod_{f=1}^3(a_1-m_f)}{(a_1-a_2)(a_1-a_3)}q_1 + \frac{\prod_{f=1}^{3}(a_1+\tilde{m}_f)}{(a_1-a_2)(a_1-a_3)}\frac{q}{q_1} + \cdots
\end{equation}
This solution agrees with \eqref{eq:simplesu3sol} for massless case. We checked the matching up to third order.

\subsection{Orbifolded instanton partition function}

The instanton partition function of SYM can be obtained from the characters at fixed points of the tangent space of the moduli space under $U(1)^2\times U(1)^N$, whose equivariant parameters are $\epsilon_i$ and $a_i$. It was observed that the moduli space of instantons in the presence of a surface defect is equivalent to the moduli space of instantons on an orbifold $\mathbb{C}\times(\mathbb{C/Z}_M)$, where $M$ is the integer characterizing the defect as in \eqref{levi}. 

For a full surface operator ($M=N$), the character of a hypermultiplet of mass $\mu$ in the bifundamental representation of $U(N)\times U(N)$ with a set of Young diagrams $\lambda=(\lambda^1,\lambda^2,...)$ and $\xi=(\xi^1,\xi^2,...)$ can be written \cite{Alday:2010vg,Kozcaz:2010yp} 

\begin{eqnarray}
&& \chi_{\mathrm{bif}}(a,\tilde{a},\lambda,\xi,\mu) = \nonumber\\
&& e^{-\mu}\sum_{k=1}^{N} \sum_{l'\geq1}e^{a_k-\tilde{a}_{k-l'}} e^{-\etwo \left(\floor{\frac{l'-k}{N}}-\floor{-\frac{k}{N}}\right)}\sum_{s=1}^{\xi_{l'}^{k-l'}}e^{-\eone s} \nonumber\\
&& - e^{-\mu}\sum_{k=1}^{N}\sum_{l\geq 1}\sum_{l'\geq 1} e^{a_{k-l+1}-\tilde{a}_{k-l'}}e^{-\etwo \left(\floor{\frac{l'-k}{N}}-\floor{\frac{l-k-1}{N}}\right)}\left(e^{-\eone \xi_{l'}^{k-l'}}-1\right)\sum_{s=1}^{\lambda_l^{k-l+1}}e^{-\eone(s-\lambda_l^{k-l+1})} \nonumber\\
&& + e^{-\mu}\sum_{k=1}^{N}\sum_{l\geq 1}\sum_{l'\geq 1} e^{a_{k-l+1}-\tilde{a}_{k-l'+1}}e^{-\etwo \left(\floor{\frac{l'-k-1}{N}}-\floor{\frac{l-k-1}{N}}\right)}\left(e^{-\eone \xi_{l'}^{k-l'+1}}-1\right)\sum_{s=1}^{\lambda_l^{k-l+1}}e^{-\eone(s-\lambda_l^{k-l+1})} \nonumber\\ 
&& + e^{-\mu}\sum_{k=1}^{N}\sum_{l\geq 1} e^{a_{k-l+1}-\tilde{a}_{k}}e^{-\etwo \left(\floor{\frac{-k}{N}}-\floor{\frac{l-k-1}{N}}\right)}\sum_{s=1}^{\lambda_l^{k-l+1}}e^{-\eone(s-\lambda_l^{k-l+1})} \,,
\end{eqnarray} 
where $\floor{x}$ is the floor function, which denotes the largest integer smaller than $x$. In this orbifold construction, the Young diagrams and adjoint scalars are periodically identified: $\lambda^{i+N} \equiv \lambda^i$, $a_{i+N}\equiv a_i$ and similarly for $\xi^i$ and $\tilde{a}_i$.

From the above character, we obtain the characters of vector multiplet and $N$ hypermultiplets of masses $M=(M_1,M_2,...,M_N)$ and $\tilde{M}=(\tilde{M}_1,\tilde{M}_2,...,\tilde{M}_N)$ in the fundamental and antifundamental representation,
\begin{eqnarray}
\chi_{\mathrm{vec}}(a, \lambda) &=& -\chi_{\mathrm{bif}}(a,a,\lambda,\lambda,0) \,,\nonumber\\
\chi_{N\mathrm{fund}}(a,M, \lambda) &=& \chi_{\mathrm{bif}}(a,M,\lambda,\emptyset,0) \,,\nonumber\\
\chi_{N\mathrm{afund}}(a,\tilde{M}, \xi) &=& \chi_{\mathrm{bif}}(-\tilde{M},a,\emptyset, \xi,0) \nonumber \,,
\end{eqnarray}
where $M$ and $\tilde{M}$ are periodically identified as well. 

The characters take the form  
\begin{equation}\label{character}
\chi(\lambda) = \sum_i s_ie^{w_i(\lambda)} \,,
\end{equation}
where $s_i$ is $\pm 1$.
The instanton partition function of a single $SU(N)$ gauge group in the presence of a full surface operator is computed in the form of
\begin{equation}
Z = \sum_{\lambda} y_1^{k_i}\cdots y_N^{k_N}\, Z_{k_1,...k_N}(\lambda),
\end{equation}
where $y_i$ is a ramified instanton counting parameter and the instanton number $k_i$ is given by
\begin{equation}
k_i = \sum_{j \geq 1}\lambda^{i-j+1}_j \,,
\end{equation}
and $Z_{k_1,...,k_N}(\lambda)$ is the product of weights $w_i$ in the numerator for $s_i=1$ and in the denominator for $s_i=-1$.

The ramified instanton partition function of $SU(3)$[1,1,1] surface operator in $SU(3)$ theory with $N_f=6$ flavors are computed
\begin{eqnarray}
Z[1,1,1] &=& 1- y_1\frac{(a_1-M_1)(a_1+\tilde{M}_2+\eone)}{\eone(a_1-a_2+\eone)}- y_2\frac{(a_2-M_2)(a_2+\tilde{M}_3+\eone)}{\eone(a_2-a_3+\eone)} \nonumber\\
&& - y_3\frac{(a_3-M_3)(a_3+\tilde{M}_1+\eone+\etwo)}{\eone(a_3-a_1+\eone+\etwo)} +\cdots
\end{eqnarray} 
We are going to impose a constraint $\sum M_i=\sum \tilde{M}_i=0$ on mass parameters. Also, we impose on $y_i$
\begin{equation}\label{y}
y_3 = \frac{y}{y_1y_2},
\end{equation}
where $y$ is the counting parameter in the absence of surface operator.
 
The twisted superpotential is extracted from the partition function and then its derivatives with respect to $y_1$ and $y_2$ are found
\begin{eqnarray}	
y_1\frac{d\cW}{dy_1} &=& - y_1\frac{(a_1-M_1)(a_1+\tilde{M}_2)}{a_1-a_2} - \frac{y}{y_1y_2}\frac{(a_3-M_3)(a_3+\tilde{M}_1)}{a_1-a_3} + \cdots \nonumber\\
y_2\frac{d\cW}{dy_2} &=& - y_2\frac{(a_2-M_2)(a_2+\tilde{M}_3)}{a_2-a_3} - \frac{y}{y_1y_2}\frac{(a_3-M_3)(a_3+\tilde{M}_1)}{a_1-a_3} + \cdots
\end{eqnarray}
If the parameters are mapped for $i=1,2$
\begin{equation}\label{paramap}
y_i = -q_i\,, \quad M_i=m_i\,, \quad \tilde{M}_{i+1}=\tilde{m}_{i}
\end{equation}
and the limit $y\,M_3\tilde{M}_1 \rightarrow -\Lambda^2$ is taken to have $N_f=4$ hypermultiplets, we get the same solution as in \eqref{quiversu3sig} and \eqref{quiversu3sig2}. This agreement is checked up to second order.

\subsection{Twisted superpotential from 5d instanton partition function}

Apart from Chern-Simons contribution, the 5d instanton partition function in the presence of a surface operator can be obtained from the corresponding 4d case by replacing each factor to the hyperbolic sine function \cite{Nekrasov:2002qd,Nekrasov:2003rj,Hollowood:2003cv,Tachikawa:2004ur}. For example, the factors for the first gauge group in \eqref{nekhiggs} are converted to
\begin{equation}
\frac{2\sinh\left(\frac{\beta}{2}(E(a_m-b_n,Y_m,W_n,s)-\mu)\right)\prod_{f=1}^{N}2\sinh\left(\frac{\beta}{2} F(a_m,Y_m,s,m_f)\right)}{2\sinh\left( \frac{\beta}{2}E(a_m-b_n,Y_m,Y_n,s) \right) 2\sinh\left( \frac{\beta}{2}(E(a_m-b_n,Y_m,Y_n,s)+\epsilon) \right)} \,.
\end{equation}

The Higgsing prescription \eqref{higgscond} applies to the five dimensional cases as well. Thus, in the case of $SU(2)[1,1]$ operator in 5d $SU(2)$ theory with $N_f=4$, we find that
\begin{equation}
Z_{5d}[1,1] = 1+ z\,\frac{\sinh\frac{\beta(a-m_1)}{2}\sinh\frac{\beta(a-m_2)}{2}}{\sinh\frac{\beta \eone}{2}\sinh\frac{\beta (2a+\eone+\etwo)}{2}} + \tilde{z}\,\frac{\sinh\frac{\beta(a+\tilde{m}_1)}{2}\sinh\frac{\beta(a+\tilde{m}_2)}{2}}{\sinh\frac{\beta \eone}{2}\sinh\frac{\beta (-2a+\eone)}{2}} + \cdots
\end{equation}
This result gives the twisted superpotential whose derivative with respect to $q_1$ agrees with the corresponding sigma model result of \eqref{5dsu2sol}.

Similarly, the partition function of $SU(3)[1,2]$ operator in 5d $SU(3)$ theory with $N_f=6$ is just the four dimensional expression \eqref{neksu3} with each factor $f$ replaced by $2\sinh(\beta f/2)$. It gives the twisted superpotential, which agrees with the sigma model result in \eqref{5dsu3sol}.

In the orbifold construction, the 5d instanton partition function is given by the product of $2\sinh(\beta w_i/2)$, where $w_i$ is a weight in the character \eqref{character}.
Then for a full surface defect in 5d $SU(3)$ theory with $N_f=4$, the derivatives of the twisted superpotential are computed 
\begin{eqnarray}
y_1\frac{d\cW_{5d}}{dy_1} &=& -y_1\frac{2\sinh\frac{\beta(a_1-M_1)}{2}\sinh\frac{\beta(a_1+\tilde{M}_2)}{2}}{\beta\sinh\frac{\beta(a_1-a_2)}{2}} - \frac{y}{y_1y_2}\frac{2\sinh\frac{\beta(a_3-M_3)}{2}\sinh\frac{\beta(a_3+\tilde{M}_1)}{2}}{\beta\sinh\frac{\beta(a_3-a_1)}{2}} +\cdots\,,\nonumber\\
y_2\frac{d\cW_{5d}}{dy_2} &=& -y_2\frac{2\sinh\frac{\beta(a_2-M_2)}{2}\sinh\frac{\beta(a_2+\tilde{M}_3)}{2}}{\beta\sinh\frac{\beta(a_2-a_3)}{2}} - \frac{y}{y_1y_2}\frac{2\sinh\frac{\beta(a_3-M_3)}{2}\sinh\frac{\beta(a_3+\tilde{M}_1)}{2}}{\beta\sinh\frac{\beta(a_1-a_3)}{2}} +\cdots\,.\nonumber
\end{eqnarray}
Parameters are mapped as in \eqref{y} and \eqref{paramap}.
If large $|M_3|$ and $|\tilde{M}_1|$ limit is taken such that
\begin{equation}
y\left(2\sinh\frac{\beta(a_3-M_3)}{2}\right)\left(2\sinh\frac{\beta(a_3+\tilde{M}_1)}{2}\right) \longrightarrow (\beta\Lambda)^2 \,,
\end{equation}
one finds the identical result as in the 3d sigma model vacuum \eqref{eq:5dsu3quiversol}. In all of the cases above, agreements are confirmed up to second order.

\section{Duality of theories for surface defect}\label{sec:duality}

\begin{figure}[ht]
	\begin{center}
		\begin{tikzpicture}[decoration={markings, mark=at position 0.6 with {\draw (-4pt,-4pt) -- (0pt,0pt);       
				\draw (-4pt,4pt) -- (0pt,0pt);}}]
		\matrix[row sep=8mm,column sep=7mm] {
			\node(lflav)[flavor]{\Large $3$} ; &&\\
			\node(gfN)[rotate=90,gaugedflavor]{\rotatebox[]{-90}{\Large $3$}};&&\node(glast)[gauge] {\Large $2$};\\
			\node(rflav)[flavor]{\Large $3$} ; &&\\
		};
		\graph{(gfN)--[postaction={decorate}](glast); (rflav)--[postaction={decorate}](gfN); (gfN)--[postaction={decorate}](lflav);(glast)--[postaction={decorate}](lflav)};
		\end{tikzpicture}
	\end{center}
	\vspace{-0.5cm}
	\caption{$SU(2)$ dual to the simple defect in $SU(3)$ gauge theory with $N_f=6$. }
	\label{dual}
\end{figure}
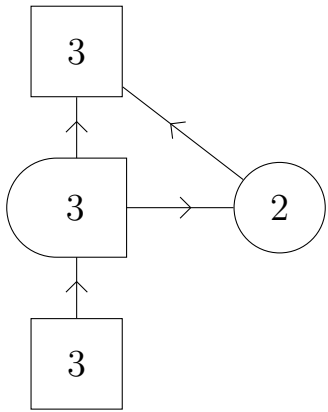

The 2d quiver theory for the surface operator of type $SU(N)[n_1,n_2,...n_M]$ with $n_I\leq n_{I+1}$ coupled to pure SYM is dual to a quiver theory, whose rank $\tilde{r}_I$ of node $I$ is given by 
\begin{equation}
\tilde{r}_I = N-r_I = \sum_{J=I+1}^Mn_J \,,
\end{equation}
with the arrow of bifundamental connecting two nodes reversed. In the dual theory, node one with rank $\tilde{r}_1$ is coupled to the four-dimensional gauge group and $\tilde{r}_I > \tilde{r}_{I+1}$ \cite{Ashok:2017lko}.   
For example, the $U(1)$ theory for $SU(3)[1,2]$ surface operator in SYM has a dual description which is $SU(2)$ theory coupled to SYM. In this section, we observe that this duality extends to the case where the bulk theory has hypermultiplets. Fig.\,\ref{dual} shows the quiver diagram of the dual $SU(2)$ theory.

The effective twisted superpotential of the dual $SU(2)$ theory with three chirals of +1 charge and three chirals of -1 charge can be written
\begin{equation}
W_d = 2\pi i\tau_{d}\sum_{i=1}^2\sigma_{di} + \sum_{i=1}^2\sum_{f=1}^3w(-\sigma_{di}+m_f) - \sum_{i=1}^2 \Tr \Big\langle w(-\sigma_{di}+\Phi) \Big\rangle \,, 
\end{equation}
where $\tau_d$ and $(\sigma_d)_{i}$ are the FI parameter and the diagonal component of adjoint scalar in 2d vector multiplet, respectively. 

A classical vacuum is chosen as
\begin{equation}
\sigma_{di} = (a_2,a_3) \,,
\end{equation}
so that $\tr\sigma_{d} = -a_1=-\sigma$ at leading order. Thus, the dual twisted superpotential at the classical level is related to the twisted superpotential of the original theory through
\begin{equation}\label{dualrel}
q_d\frac{dW_d}{dq_d} = -q_1\frac{dW}{dq_1} \,,
\end{equation}
where $q_d=e^{2\pi i\tau_d}$ and $W$ is from \eqref{eq:simplesu3sol}. One can set $\tau_d = -\tau_1$ of the original theory so that $W_d = W$. This relation can be checked at higher order by solving the chiral ring equation. 

The twisted chiral ring relations of the dual theory are given by
\begin{equation}
(1+q)P_2(\sigma_{di}) = q_1\prod_{f=1}^3(\sigma_{di}-m_f) + \frac{q}{q_1}\prod_{f=1}^3(\sigma_{di}+\tilde{m}_f) \,,
\end{equation} 
for $i=1,2$. We computed $\sum_i(\sigma_{d*})_{i}$ up to second order for massive case and up to fourth order for massless case and observed that it is minus of $\sigma_*$ in \eqref{eq:simplesu3sol}, confirming the duality relation \eqref{dualrel}. 

The duality can also be seen from the instanton partition function by dual Higgsing prescription. The twisted superpotential obtained via Higgsing prescription of \eqref{higgssu3} can also be reproduced by 
\begin{equation}\label{dualhiggs}
\vec{b}=\vec{a}+(0,\etwo,\etwo) \qquad,\qquad \mu=0 \,.
\end{equation} 
The resulting partition function looks different from the original one in \eqref{neksu3} but it yields the same twisted superpotential of the original theory with a minus sign.

In 3d/5d coupled systems, the Chern-Simons level must be determined for the duality relation. In the example we are considering, that is an $SU(3)$[1,2] defect in $SU(3)$ theory with $N_f=6$ hypermultiplets, we found that the duality holds when the CS level is zero. This fits nicely with the fact that the CS level of the defect in SYM was found to be $\pm1/2$ \cite{Ashok:2017lko}. Starting from the theory with zero CS level, one can integrate out the matter chiral fields and generate CS contributions \cite{Redlich:1983dv,Redlich:1983kn} for a simple surface defect in 5d SYM. Then eq.\eqref{cs} indicates that the CS level should be $\pm1/2$ for $SU(3)$ SYM.

The chiral ring equations are
\begin{equation}
(1+q)\widehat{P}_2(\sigma_{di}) = q_1\prod_{f=1}^3 \left(2\sinh\frac{\beta(\sigma_{di}-m_f)}{2}\right) + \frac{q}{q_1}\prod_{f=1}^3 \left(2\sinh\frac{\beta(\sigma_{di}+\tilde{m}_f)}{2} \right) \,.
\end{equation}
The vacuum solution $\sum_i(\sigma_{d*})_{i}$ is found to be minus of $\sigma_*$ given in \eqref{5dsu3sol}. 

Applying the dual Higgsing prescription \eqref{dualhiggs} to the 5d instanton partition function, one can obtain the twisted superpotential, which reproduces the negative of vacuum solution to the original chiral ring equation. 

In \cite{Ashok:2017lko, Ashok:2017bld, Ashok:2018zxp}, the duality have been checked using the contour integral expression of the instanton partition function, where it was interpreted as a different choice of contour prescriptions.

\section{Discussion}
\label{sec:dis}

We have shown that the vacua of 2d (2,2) sigma models coupled to 4d SQCD such that the sum of matter charges is zero correspond to the logarithmic derivative of the twisted superpotential extracted from the instanton partition function of the bulk theory in the presence of a surface operator. The exponential of 2d FI parameter is identified with the ramified instanton counting parameter such that the product of them amounts to the exponential of 4d gauge coupling constant. We have checked that the Seiberg-like dual theory for the sigma model reproduces the same vacuum solution to the chiral ring. The twisted superpotential of the dual theory are also obtained from the instanton partition function by the dual Higgsing prescription.
We have performed similar analysis for 3d/5d coupled systems. The duality holds for a specific value of the CS level. In the example we have considered the CS term vanished. 

One could study more general types of surface operators along the line of this paper. The equivariant character for the instanton partition function in the presence of a generic surface defect is known in the literature \cite{Wyllard:2010vi, Kanno:2011fw}. In fact, using the character, we have computed the instanton partition function for some $SU(4)$ theories in the presence of generic surface operators and found agreement with 2d sigma model results, which was not reported in the work. 

Another way to compute the instanton partition function in the presence of a surface defect is to use the two-sphere partition function of a suitable (2,2) sigma model by considering the resolution of the orbifold \cite{Bonelli:2013rja}, where D1 branes are wrapped on the blown up sphere. Other directions of research would include the study of surface defects in four or five dimensional quiver theories \cite{Nekrasov:2013xda}, in $\cN=1$ theories \cite{Gaiotto:2013sma}, and in the context of integrable systems, where the twisted chiral ring relations correspond to the Bethe equations. \cite{Nekrasov:2017gzb,Nekrasov:2017rqy, Sciarappa:2017hds,Jeong:2018qpc}

\begin{appendix}
\section{Calculation of quantum vevs}
\label{sec:app}

\subsection{4d theories}\label{qv4d}
In $SU(N)$ gauge theory with $N_f=2N$, the SW curve can be written 
\begin{equation}
y^2(x) = (1+q)^2P_N^2(x) - 4qB_{N_f}(x) 
\end{equation}
and the SW differential $\lambda(x)$ is given by
\begin{equation}
\lambda(x) = x d\log((1+q)P_N(x)+y(x)) \,,
\end{equation}
where $P_N(x)$ and $B_{N_f}(x)$ are the characteristic polynomials \eqref{PN} and \eqref{BN}.

The period integral 
\begin{equation}
a_i = \frac{1}{2\pi i}\oint_{e_i}\lambda(x)
\end{equation}
can be computed by expanding $\lambda(x)$ in small $q$. The residue computation can be simplified using the integration by parts and dropping total derivatives. After finding residues, $e_i$ can be solved in terms of $\vec{a}$. They satisfy $\sum_ie_i=0$. 

In $SU(2)$ gauge theory with $N_f=4$, the quantum vevs are  
\begin{eqnarray}
e_1 &=& a -\frac{q}{4a^3}(3a^4+(m_1m_2+\tilde{m}_1\tilde{m}_2)a^2-m_1m_2\tilde{m}_1\tilde{m}_2) + \cO(q^2) \,,\nonumber\\
e_2 &=& -e_1\,, \nonumber
\end{eqnarray}
where $\sum_ia_i=\sum_im_i=\sum\tilde{m}_i=0$ have been used.  

In $SU(3)$ gauge theory with $N_f=6$, they are found for massless case 
\begin{eqnarray}
e_1 &=& a_1 - q\,\frac{2 a_1^5 (a_1^2 + 3 a_2a_3 - 2 a_1(a_2 + a_3)}{(a_1 -a_2)^3 (a_1 - a_3)^3}+ \cO(q^2) \,,\nonumber\\
e_2 &=& a_2 - q\,\frac{2 a_2^5 (a_2^2 + 3 a_1a_3 - 2 a_2(a_1 + a_3)}{(a_2 -a_1)^3 (a_2 - a_3)^3}+ \cO(q^2) \,,\nonumber\\
e_3 &=& a_3 - q\,\frac{2 a_3^5 (a_3^2 + 3 a_1a_2 - 2 a_3(a_1 + a_2)}{(a_3 -a_1)^3 (a_3 - a_2)^3}+ \cO(q^2) \,\nonumber
\end{eqnarray} 

For $SU(N)$ theory with $N_f\leq 2N$, the curve is 
\begin{equation}
y(x)^2 = P_N(x)^2 -4\Lambda^{2N-N_f}B_{N_f} \,,
\end{equation} 
with differential $\lambda = xd\log(P_N(x)+y(x))$. 
In the case of $N=3$ and $N_f=4$, $e_i's$ are given by
\begin{eqnarray}
e_1 &=& a_1 - \Lambda^2\,\frac{2 a_1^3 (2 a_2a_3 -  a_1(a_2 + a_3))}{(a_1 -a_2)^3 (a_1 - a_3)^3}+ \cO(\Lambda^4) \,,\nonumber\\
e_2 &=& a_2 - \Lambda^2\,\frac{2 a_2^3 ( 2a_1a_3 -  a_2(a_1 + a_3))}{(a_2 -a_1)^3 (a_2 - a_3)^3}+ \cO(\Lambda^4) \,,\nonumber\\
e_3 &=& a_3 - \Lambda^2\,\frac{2 a_3^3 (2 a_1a_2 -  a_3(a_1 + a_2))}{(a_3 -a_1)^3 (a_3 - a_2)^3}+ \cO(\Lambda^4) \,,\nonumber
\end{eqnarray}
for massless flavors. 

\subsection{5d theories}\label{qv5d}

The SW curve in 5d SQCD for $N_f=2N$ can be written 
\begin{equation}
Y^2(x) = (1+q)^2\widehat{P}_N^2(x) - 4q\widehat{B}_{N_f}(x) \,,
\end{equation}
with the 5d differential
\begin{equation}
\widehat{\lambda} = xd\log((1+q)\widehat{P}_N(x) + Y(x)) \,,
\end{equation}
where $\widehat{P}_N(x)$ and $\widehat{B}_{N_f}(x)$ are defined in \eqref{5dPN}.
 We first perform the integral 
\begin{equation}
 a_i = \frac{1}{2\pi i}\oint_{\widehat{e}_i}\widehat{\lambda}(x)\,.
\end{equation}
For 5d $SU(2)$ theory with $N_f=4$, 
\begin{equation}
\widehat{e}_1 = a -\frac{2q}{\beta\sinh^2\beta a}\left(\sum_{i=1}^4\cosh\frac{\beta(a-m_i)}{2}\prod_{j\ne i}^4 \sinh\frac{\beta(a-m_j)}{2}  -\coth\beta a \prod_{i=1}^4\sinh\frac{\beta(a-m_i)}{2}\right) +\cO(q^2)\nonumber
\end{equation}
with $\widehat{e}_2=-\widehat{e}_1$. We have used the notation $m_3=-\tilde{m}_1$, $m_4=-\tilde{m}_2$ and $\sum_{i=1}^4m_i=0$.

For 5d $SU(3)$ theory with $N_f=6$ massless hypermultiplets and zero Chern-Simons level, the quantum vev's are given by
\begin{eqnarray}
\widehat{e}_1 &=& a_1 +q\frac{4\sinh^6(\frac{\beta}{2}a_1)}{\beta\sinh^2(\frac{\beta }{2}a_{12})\sinh^2(\frac{\beta}{2}a_{13})}\left(-3\coth(\frac{\beta}{2}a_1)+\coth(\frac{\beta}{2}a_{12}) +\coth(\frac{\beta}{2}a_{13}) \right) +\cO(q^2)\,,\nonumber\\
\widehat{e}_2 &=& a_1 +q\frac{4\sinh^6(\frac{\beta}{2}a_2)}{\beta\sinh^2(\frac{\beta }{2}a_{21})\sinh^2(\frac{\beta}{2}a_{23})}\left(-3\coth(\frac{\beta}{2}a_2)+\coth(\frac{\beta}{2}a_{21}) +\coth(\frac{\beta}{2}a_{23}) \right) +\cO(q^2)\,,\nonumber\\
\widehat{e}_3 &=& a_1 +q\frac{4\sinh^6(\frac{\beta}{2}a_3)}{\beta\sinh^2(\frac{\beta }{2}a_{31})\sinh^2(\frac{\beta}{2}a_{32})}\left(-3\coth(\frac{\beta}{2}a_3)+\coth(\frac{\beta}{2}a_{31}) +\coth(\frac{\beta}{2}a_{32}) \right) +\cO(q^2)\,,\nonumber
\end{eqnarray}
where $a_{ij}=a_i-a_j$. 

For 5d $SU(3)$ theory with $N_f=4$ massless hypermultiplets, we obtain
\begin{eqnarray}
\widehat{e}_1 &=& a_1 +\frac{(\beta\Lambda)^2\sinh^3(\frac{\beta}{2}a_1)}{2\beta\sinh^3(\frac{\beta}{2}a_{12})\sinh^3(\frac{\beta}{2}a_{13})}\left(-2\cosh(\frac{\beta}{2}(a_1-a_2-a_3))+\cosh(\frac{\beta}{2}(a_{1}+a_2-a_3)) \right. \nonumber\\
&& \hskip 9cm \left. +\cosh(\frac{\beta}{2}(a_{1}-a_2+a_3)) \right) +\cO(q^2)\,,\nonumber\\
\widehat{e}_2 &=& a_2 +\frac{(\beta\Lambda)^2\sinh^3(\frac{\beta}{2}a_2)}{2\beta\sinh^3(\frac{\beta}{2}a_{21})\sinh^3(\frac{\beta}{2}a_{23})}\left(-2\cosh(\frac{\beta}{2}(a_2-a_1-a_3))+\cosh(\frac{\beta}{2}(a_{2}+a_1-a_3)) \right. \nonumber\\
&& \hskip 9cm \left. +\cosh(\frac{\beta}{2}(a_{2}-a_1+a_3)) \right) +\cO(q^2)\,,\nonumber\\
\widehat{e}_3 &=& a_3 +\frac{(\beta\Lambda)^2\sinh^3(\frac{\beta}{2}a_3)}{2\beta\sinh^3(\frac{\beta}{2}a_{31})\sinh^3(\frac{\beta}{2}a_{32})}\left(-2\cosh(\frac{\beta}{2}(a_3-a_1-a_2))+\cosh(\frac{\beta}{2}(a_{3}+a_1-a_2)) \right. \nonumber\\
&& \hskip 9cm \left. +\cosh(\frac{\beta}{2}(a_{3}-a_1+a_2)) \right) +\cO(q^2)\,.\nonumber
\end{eqnarray}

\end{appendix}

\providecommand{\href}[2]{#2}\begingroup\raggedright
\endgroup

\providecommand{\href}[2]{#2}\begingroup\raggedright
\bibliography{references}
\bibliographystyle{JHEP}
\endgroup

\end{document}